\documentclass[twocolumn,openany,openright,amsmath,amssymb,prb ]{revtex4}

\usepackage{float}
\usepackage{latexsym} 
\usepackage{amsmath,amsthm}
\usepackage{ifpdf}
\usepackage{epstopdf}

\usepackage{dcolumn}
\usepackage{bm}
\usepackage{braket}

\usepackage{color}

\usepackage{graphicx}

\usepackage[final]{changes}

\newcommand{\hatF}{\hat{\phi}}%
\newcommand{\kk}{\mathbf{k}}

\newcommand{\rr}{\mathbf{r}}

\newcommand  {\opa}[1] {\hat a #1}
\newcommand  {\opb}[1] {\hat b #1}
\newcommand  {\ad}[1] {\hat a^{\dagger} #1}
\newcommand  {\bd}[1] {\hat b^{\dagger} #1}

\begin{document}
\def \k{\bold k}
\def \l{\bold l}
\def \i{\bold i}
\def \p{\bold p}
\def \r{\bold r}
\def \q{\bold q}
\def \A{\bold A}
\def \beq{\begin{equation}}
\def \eeq{\end{equation}}
\def \beal{\begin{aligned}}
\def \eal{\end{aligned}}
\def \bes{\begin{split}}
\def \ees{\end{split}}
\def \besu{\begin{subequations}}
\def \esu{\end{subequations}}
\def \g{\gamma}
\def \G{\Gamma}
\def \ac{\alpha_c}

\title {Long-lived Nonlinear Oscillatory States  in Interacting Relativistic Bose-Einstein Condensates}
\author{Leone Di Mauro Villari}
\author{Ian Galbraith}
\author{Fabio Biancalana}
\affiliation{Institute of Photonics and Quantum Sciences, School of Engineering and Physical Sciences, SUPA, Heriot-Watt University, Edinburgh EH14 4AS, UK.}

\begin{abstract}
We study a mean field model for the dynamics of an interacting Bose-Einstein condensate in two dimensional pseudo-relativistic materials. This model is relatively simple,  but contains long-lived solutions called oscillons which are absent in simple non-relativistic condensates. We report on a variety of scenarios including interactions between pairs of oscillons and oscillons propagating across an inhomogeneous material boundary. Hitherto relativistic oscillons  have been studied only in high energy physics and cosmology and their relevance has not been highlighted so far in condensed matter physics. 
\end{abstract}
\maketitle

\section{introduction}
 Many decades after the theoretical prediction of Bose and Einstein, Bose-Einstein condensates (BECs) were  experimentally detected in laser-cooled, magnetically-trapped ultracold bosonic atomic clouds \cite{bs,ae,gl}. 
More recently, BECs have also been seen  in fermionic atomic gases as a result of fermions pairing into bosons \cite{lw}. An interesting and widely studied example of fermions paring are the excitonic bound states of electron and holes in semiconductors. The possibility for semiconductor excitons to undergo Bose-Einstein condensation has been suggested long ago, at the beginning of the sixties \cite{bb}. 
As the critical temperature for elementary boson condensation scales as the inverse of the boson mass, it was thought that exciton condensation could be obtained at 100\,K or even at room temperature, the exciton mass being much less than the free electron mass. 
The experimental search for  the condensed phase, however, turned out to be challenging mostly due to the fact that excitons are not an elementary but a composite boson with a finite lifetime\cite{cc}. Despite these difficulties signatures of exciton condensation have been reported in double quantum wells \cite{se,bg,bl,nf}, microcavities \cite{bh}, graphene \cite{lw} and transition metal dichalcogenoides \cite{ko}.  It is known that excitons and exciton-polaritons show a BEC-like insulating phase, that has  been the subject of promising theoretical and experimental investigation mainly in graphene-like real \cite{sg,rn,sk} and synthetic \cite{k} lattices as well as in topological insulators \cite{wh}. This BEC-like phase is particularly interesting from an experimental and theoretical point of view since it presents resents the crossover behaviour from the Bardeen-Cooper-Schrieffer (BCS) limit to the Bose-Einstein condensation limit. Recently exciton condensates with superfluids transport properties have been observed in double bilayer graphene \cite{lt} and Van der Waals heterobilayers  \cite{zt}. Due to the pseudo-relativistic behavior of low energy quasiparticles in a honeycomb lattice one may wonder what are the relevant properties  of the dynamics in the condensed phase. 
Moreover the relevance of relativistic BECs has been recently pointed out in gases with both electron and hole pairing. Relativity comes into play as those two composite bosons form a particle antiparticle pair \cite{gl}.  The boson-boson interaction in relativistic BECs could be potentially used to experimentally mimic field theory in condensed matter. It is also a promising system for analogue model of gravity \cite{ff}.

In this paper, starting from the dispersion of excitons in Dirac-like materials, we derive and investigate a simple, but flexible mean field model that can describe the dynamics of the condensed phase in different physical scenarios. Central to this model is a relativistic generalisation of the Gross-Pitaevskii  equation (GPE), i.e a nonlinear Klein-Gordon equation.
In Section II we derive the exciton dispersion relation, using a two particle model. In section III we study the condensate phase of the exciton and derive the mean field model. In Section IV we investigate the properties of a non-stationary, but localised solution of the model, known as an oscillon. In Sections V and VI we study the oscillon dynamics in more complicated scenarios: when they interact with each other  and  with a localized defect,  and, when two materials with different energy gap are in contact. Until now the oscillon solutions of relativistic field theories have been studied only in high energy physics and cosmology, in particular for self-interacting scalar fields and in the $SU(2)$ Higgs model \cite{fg}. Oscillon solutions exist also in non-relativistic BECs, but their mathematical structure is different.  In such systems, in order to stabilise oscillating solutions, one needs either consider coupled equations, or apply some sort of external perturbation.  A discussion of non-relativistic oscillon solutions in BECs can be found in \cite{ss,rv}.  In these papers the authors consider coupled BECs  \cite{ss}  and a system confined in a trap with oscillating walls \cite{rv}.  Oscillons in nonlinear and parametrically driven systems have been studied and experimentally observed in fluids such us  granular media \cite{um}, Newtonian fluids \cite{af} and colloidal suspensions \cite{lh}. The formation and interaction of oscillons have been the subject of several theoretical, numerical, and experimental studies. However, a number of open problems remains, mostly related to their stability and to the transition between radiative and non-radiative oscillons. Recent papers studying parametrically driven systems in one dimension focussed on the transition between localised structures to breathing localised states \cite{bz,bz1}. In two dimensions oscillon instabilities  have been numerically studied in magnetic systems \cite{lc} and  in the  parametrically  driven damped  sine-Gordon equation \cite{ab}.
\section{Exciton dispersion in Dirac Materials}
The Hamiltonian describing the low energy behavior of Dirac quasi-particles around one $K$ point of the reciprocal honeycomb lattice reads
\beq \label{eq1}
H_1 = \left(\begin{array}{cc}\Delta/2 & v p e^{i \theta} \\v p e^{-i\theta} & -\Delta/2\end{array}\right),
\eeq
where $\Delta$ is the energy gap and $v$ the Fermi velocity, $p=\hbar k$ the modulus of the electron momentum and $\theta = \arctan(p_y/p_x)$. Such dispersion is appropriate to describe low energy electrons in gapped graphene and, with certain limitations, transition metal dichalcogenides (TMDs).  Assuming zero center of mass momentum  for excited $e-h$ pairs, the two particle Hamiltonian, without Coulomb interaction is given by 
the tensor product $H_2 = H_1 \otimes I_2 - I_2\otimes(T H_1 T^{-1})$, with $I_2$ the $2 \times 2$ identity matrix and $T$ the time reversal operator. It reads 
\beq \label{eq2}
H_2 = \left(\begin{array}{cccc}0 & v p e^{i\theta} & v p e^{-i\theta} & 0 \\ v p e^{-i\theta} & \Delta & 0 & v p e^{-i \theta} \\v p e^{i\theta} & 0 & -\Delta & v p e^{i\theta} \\0 & v p e^{i\theta} & v p e^{-i \theta}& 0\end{array}\right),
\eeq 
and this matrix has four eigenvalues: $\pm 2\sqrt{ v^2 p^2 + \Delta^2/4}$ and a doubly degenerate zero eigenvalue \cite{rn}. The zero-energy eigenstates correspond to the cases when the system has a single electron or a hole with its complementary particle in the negative energy sea  \cite{tg,rn}.

\section{Condensate states of excitons}
To study condensed phases in a Dirac material, the Hamiltonian in equation (\ref{eq2}) needs to be modified to include Coulomb interactions \cite{rn}. They can be introduced  by deriving a set of renormalised Dirac-Bloch equation \cite{sk,cm,cb,msk,lg}. When electrons and holes have the same mass, the Hamiltonian in equation (\ref{eq2}) can be block diagonalised \cite{tg}. We set up the bands in such a way that  zero energy  is  located  half way  between  their  extrema.  This results in a reduced electron-hole Hamiltonian \cite{rn}
\beq \label{he}
H_E = \left(\begin{array}{cc}\Delta/2 & v q e^{i \theta_\q} \\v q e^{-i\theta_\q} & -\Delta/2\end{array}\right).
\eeq
where $q = |\q|$ is the modulus of the exciton's momentum in the centre of mass frame and $\theta_\q$ the related angle \cite{rn}. After introducing Coulomb interaction the eigenvalue problem for the Hamiltonian $H_E$ can be solved analytically giving the exciton energies and wave-functions \cite{nov,lg}. The energy levels of the exciton series are given by
\beq \label{eq3}
E_{n,j}  = \frac{\Delta}{\sqrt{1+ \frac{\alpha_c^2}{(n+\gamma)^2}}},
\eeq
where $n$ is the principal quantum number and $\gamma = \sqrt{j^2-\alpha^2_c}$ with $j=m+ 1/2$ being the eigenvalue of the pseudospin-angular momentum. The constant $\alpha_c$ is the dimensionless Coulomb coupling strength  and  the  spinor wavefunction is of the form
\beq \label{eq4}
 \vec \Psi_{n,j} (\bold q) = \left(\begin{array}{c}\varphi_{n,j}(\q) \\  \pm i \chi_{n,j}(\q) \end{array}\right).
\eeq
From equation (\ref{eq3}) we can observe that if the coupling constant exceeds the critical value ($\alpha_c = \frac{1}{2})$, the ground state energy becomes imaginary and a phase transition to an excitonic insulator occurs \cite{sk}.  The excitonic insulator state is a BCS-like condensate of excitons that can show a BCS-BEC crossover at low density \cite{kk}. It can be regarded as a coherent superposition of the non-interacting ground-state and all exciton states with vanishing real part of the lowest energy level $E_{0,\frac{1}{2}}$ \cite{sk}. This state is more complicated than a normal BEC since at strong Coulomb coupling the quasiparticle picture becomes less accurate and many-body theory may be needed. A description of this state at the mean field level at low density is presented in Ref. \onlinecite{we}.
In what follows we focus on the low Coulomb interaction regime $\alpha_c < 1/2$ where the use of the excitonic limit is more appropriate. In this regime the ground state for the excitons is the normal $1s$ state and the system can undergo a phase transition to a BEC state when cooled below a critical temperature $T_c$. For exciton systems this temperature can be around $100$ K or higher\cite{cc}. A detailed description of the macroscopic coherent ground state of an exciton condensate is given in chapter 2 of Ref. \onlinecite{ms}.
We shall now derive the mean field model that describes the dynamics of the condensate state.  We consider  $X^0$-type excitons only, with  spin and pseudo-spin both equal to zero.

In the low density  limit the system can be seen as a weakly interacting Bose gas of  excitons. The non-interacting, first quantised, Hamiltonian of a pseudo-relativistic gas of bosons   reads  
\beq \label{h0}
H_0 = \sqrt{ \hbar^2 v^2 \hat k^2 +  m^2 v^4},
\eeq
where $\hat k$ is the momentum operator and $m = \hbar^2 (d^2 \mathcal{E}_\k)/d k^2)^{-1}= \Delta/(4 v^2)$ is the exciton effective mass with $\mathcal{E}_\k$ being the exciton dispersion. In relativistic quantum mechanics Dirac proved the equivalence between the ill defined operator $H_0$ and Hamiltonian $H_E$ for spin-$1/2$ massive particles. The same equivalence can be established also for scalar particles by regarding the scalar wave-field as a doublet. A detailed explanation of this can be found in section four of Ref. \onlinecite{aie}.
\\To simplify our treatment we approximate the exciton-exciton interaction with a hard sphere potential \cite{L,ms},
\beq \label{eq7}
\mathcal{U}(|\rr-\rr'|)= N\frac{4\pi \hbar^2}{m L_{\mathit{eff} } }a_B \delta(\rr-\rr'),
\eeq
where  $a_B$ is the $1s$ exciton Bohr radius, $L_{\mathit{eff}}$ the effective thickness of the monolayer and $N$ the number of particles. The use of this approximation for the interaction potential is justified within the low density limit ($n_{ex} a_B^2 \ll 1$). In other words since the excitons are neutral, when the density is low enough to ignore the fermionic nature of the electron-hole pairs, one can assume that the range of interactions is of the order of the exciton Bohr radius. 

Within this approximation we can compute  the field Hamiltonian from $H_0$ and the interacting potential in equation (\ref{eq7}).  We define first the exciton creation operator, in one of the two equivalent $K$-points, using the spinor in equation (\ref{eq4})
\beq \label{eq6}
\begin{aligned}
\hat c^\dagger_{\k} &= \sum_{\q} (\varphi(\q) + i \chi(\q))  \ad_{\k+\q} \bd_{\k-\q},
\end{aligned}
\eeq
where $\{\ad_\k ,\bd_\k,\opa_\k,\opb_\k\}$ are ladder operator for electrons and holes. More details on the definition of the exciton operator are given in Appendix A. We drop the $(n,j)$ indices since we are considering condensed excitons in the ground state. The field operator for this type of pseudo-relativistic scalar excitons is written as
\beq \label{fi}
\hatF(\rr,t) = \frac{1}{\sqrt A} \sum_\k \frac{1}{\sqrt{\mathcal E_\k}} (\hat c_\k e^{-i(\k \cdot \r - \omega_\k t)} +  \hat c^\dagger_\k e^{i(\k \cdot \r - \omega_\k t)}),
\eeq
where  $A$  the area of the layer. We can now use this field operator and equations (\ref{h0}) and (\ref{eq7})  to compute the field Hamiltonian as
\beq
\begin{split}
H(\hatF)= &\int d^2r \,  \hatF(\rr) \Bigl(i \hbar \frac{\partial} {\partial t} - H_0  \Bigl )^2 \hatF(\rr)  \\ + &\int d^2 r \, d^2 r' \,  \hatF(\rr)  \hatF(\rr') \, \mathcal{U}(|\rr-\rr'|)  \, \hatF(\rr)  \hatF(\rr'),
\end{split}
\eeq
where we have squared the linear term, a standard procedure in high energy physics to derive the Klein-Gordon Hamiltonian from the pseudo-differential operator $H_0$.  After some algebra we obtain a pseudo-relativistic generalisation of the GPE \cite{ff},
\beq \label{eq7a}
H(\hatF) = \frac{1}2\int d^2r \,\Bigl[ \hbar^2 \hatF_t^2 + \hbar^2 v^2 (\nabla \hatF)^2 + m^2 v^4 \hatF^2 + 2 U_0 \hatF^4\Bigl].
\eeq 
Here  $U_0 =  N 4\pi  \hbar^2  m v^4 a_B / L_{\mathit{eff} }$, and this nonlinear coupling constant has dimensions of energy cubed $\times$  area. The Hamiltonian in equation (\ref{eq7a}) is Lorentz invariant with velocity $v$,  the Fermi velocity of the carriers. The mismatch between the Fermi velocity and the speed of light will break the Lorentz invariance when the system is coupled with an external electromagnetic field. We should expect the same symmetry breaking, at high density, beyond the validity of the excitonic limit. In this case the Fermi nature of electrons and holes is relevant and the dipole interaction would again break Lorentz invariance. Hamiltonians in equations (\ref{eq1}) and (\ref{he}) are first order in the low momentum ($\kk \cdot \p$) expansion around the $K-$point. It is worth to stress here that higher order terms, such as trigonal warping and electron-hole asymmetry, can break Lorentz invariancecite{neto}. It has been shown, however, that these distortions of the band structure are negligible in graphene and many graphene-like systems thus the Dirac quasi-particle picture is appropriate in many realistic situations. This makes the derivation of equation (\ref{eq7a})  consistent with real world experiments \cite{neto}. \\
In terms of ladder operators the Hamiltonian  in equation (\ref{eq7a}) can be written as (see Appendix B),
\beq \label{eq8a}
H = \sum_{\k} \mathcal E_{\k}  ~\hat c^\dag_\k \hat c_\k +  \frac{U_0}{2A} \sum_{\k \l \p} \frac{1}{\sqrt{\mathcal E_{\k}\mathcal E_{\l} \mathcal E_{\l + \p } \mathcal E_{\k-\p}} } \hat c^\dag_\k \hat c^\dag_\l \hat c_{\l + \p} \hat c_{\k-\p},
\eeq
where the second term is the standard four field interaction commonly used to model the condensed phase of an interacting gas of excitons in the structureless particle approximation \cite{ms}, namely when we can neglect the fermionic nature of the electron-hole pairs. The energy dependent pre-factor comes from the normalisation of the exciton scalar field.
 When the bosons are in a condensate state, it is then possible to describe the dynamics of the condensate at the mean-field level by performing the substitution $\hatF(\r,t) \to \phi(\r,t)$: the order parameter $\phi$ satisfies then the classical equation
 \beq \label{eq8}
 \square \phi  -  \mu^2 \phi - \tilde U_0 \phi^3 = 0,
 \eeq\\
with $\mu = m v/ \hbar$, $\tilde U_0 =U_0/\hbar^2 v^2 $ and we adopted the standard definition for the flat space box operator
 \beq \label{eq9}
 \square =  - \frac{1}{v^2} \partial^{2}_t + \nabla^2, 
 \eeq 
 with the Fermi velocity $v$. Equation (\ref{eq8}), in $2+1$-dimensions has non-stationary localised solutions called oscillons \cite{fg,hs,cg,hc}.
Oscillons are metastable solutions with a very long lifetime that depends critically on the initial condition \cite{hs,cg,hc}. The longevity of the oscillon lifetime has been extensively reported in many numerical studies \cite{hs,fg,hc} along with their soliton-like properties \cite{hs,cg}. In the following sections we shall review some of these properties and explore further aspects of these solutions. Before proceeding to the solution of  equation (\ref{eq8}) we shall give a brief summary on its validity. Being a generalisation of the GPE,  equation (\ref{eq8}) retains its limitations, the exciton-exciton four fields interaction is valid either in the diluted regime in consideration here, or in the opposite limit (high density) when the coupling is very weak $U_0 \approx 1/N$, where $N \gg 1$ is the number of excitons. Moreover this formulation holds when finite temperature effects are negligible. We note that a finite temperature may be included in the mean field picture with a simple modification of  equation (\ref{eq7a}) \cite{ms}.
 \section{Oscillons in relativistic BECs}
\begin{figure}
\begin{center}
\includegraphics[width=1 \columnwidth]{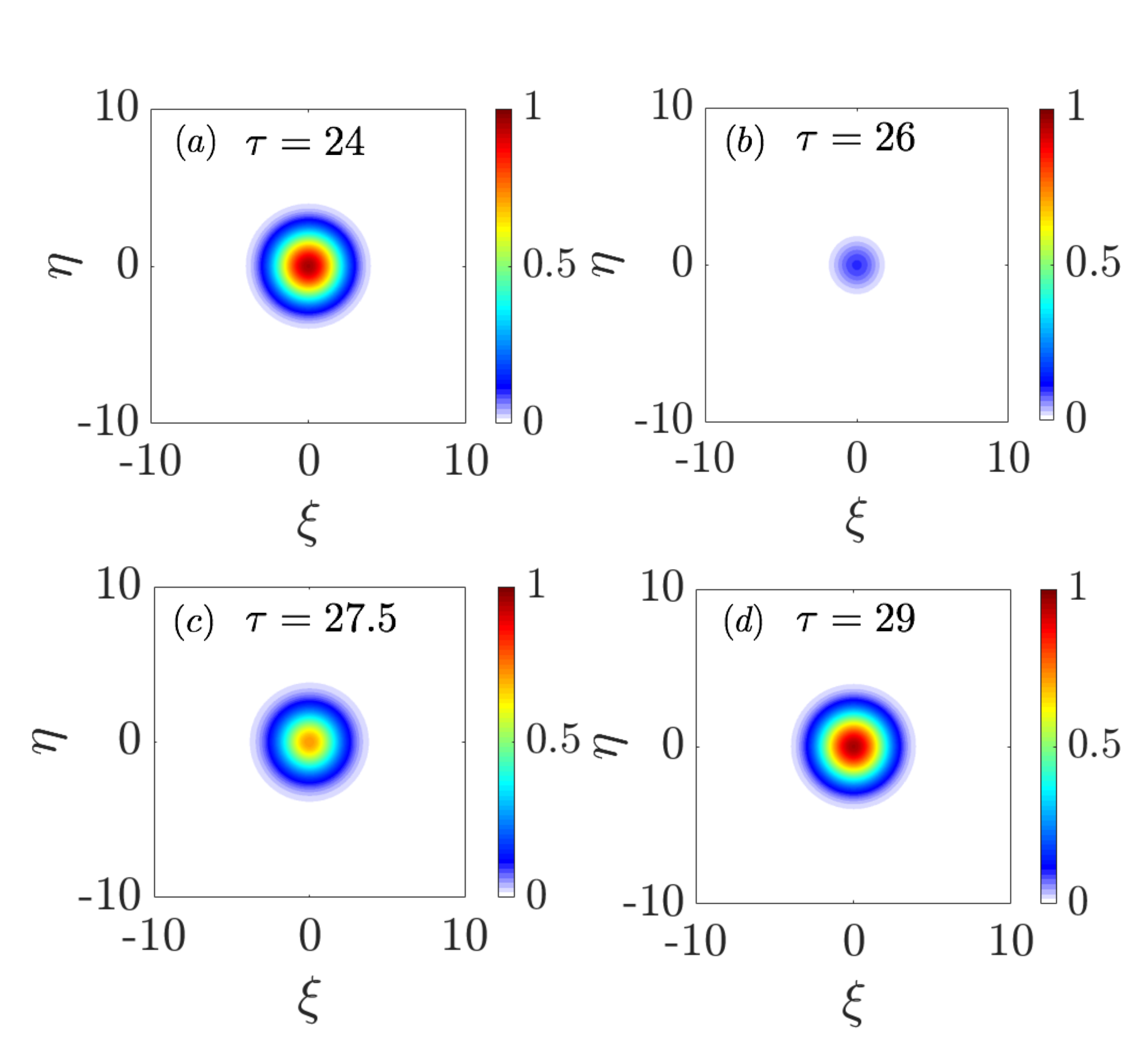}
\caption{Dynamics of the modulus square of the field, in time, after subtracting the background. (a) The initial oscillon state, (b) the first collapse, (c) and (d) the first revival. The first snapshot is at $\tau=24$ after the initial transient has quenched.}
\label{fig1}
\end{center}
\end{figure}

In this section we first review the emergence of oscillons in the NLKE in equation (\ref{eq8})  \cite{hs,cg,hc}.  We then discuss the relation between the dispersion term and the oscillon's formation. To do so we introduce a system of scaled space and time variables given by 
 \beq \label{scaling}
 \xi =\frac {x}{x_0},  \,\,  \eta =\frac{y}{y_0},  \,\,\tau = \frac{t \, \Delta} {\hbar},
 \eeq
 and define $r_0 = \sqrt{x_0^2+ y_0^2}$.  $x_0$ and $y_0$ are chosen appropriately for the initial conditions of the problem and for our purposes we always choose $x_0 = y_0$.  
 In these variables equation (\ref{eq8}) reads
 \beq \label{eq10}
 \partial^2_{\tau} \psi - \beta (\partial_\xi^2 + \partial_\eta^2)\psi + \psi + \psi^3 = 0,
 \eeq
 where $\beta =8 [ \hbar v/(r_0 \, \Delta)]^2$, and the dimensionless field $\psi(\xi,\eta,\tau)$ is defined as 
 \beq \label{eq13}
 \psi = 4 \hbar v  \sqrt{\frac{ N \pi a_B}{L_{\mathit{eff}}\Delta}} \phi.
 \eeq
Equation (\ref{eq10}) has been solved using both a pseudo-spectral implicit method and a finite difference leapfrog algorithm \cite{hs}. In what follows we present the results for a gapped graphene sample $\Delta = 0.2$\,eV, $v = c/300 $ and $\beta=1$, which implies   $r_0 = 9.3 $\,nm. Figure \ref{fig1} shows the dynamics of the modulus square of the field $\psi$ when the initial state comprises a uniform condensate  background with a gaussian shaped hole at the origin, i.e.
 \beq \label{init}
 \psi(\xi,\eta) = A_0 \left(1-  e^{-\frac{\xi^2+\eta^2}{\sigma^2}}\right),
 \eeq 
 with $A_0=1$ and $\sigma = 2.86$ which is chosen  so that the oscillon solution is maximally metastable \cite{hs,cg}, i.e. it has the maximal lifetime. This means that in this configuration the nonlinearity best compensates the dispersion.
We note that a satisfactory explanation for the metastability of oscillon solutions is still missing. There is not an obvious relation between the long lifetime of the oscillon and the symmetries of the Hamiltonian. This is due to the fact that the NLKGE is not integrable and thus there is no direct link between its solutions and conservation laws.  However we know that in non-relativistic systems localised solution with periodic oscillations are not attractors of the dynamic unless one considers more complicated systems \cite{ss, rv}, as we have mentioned in the introduction. This suggests that the pseudo-relativistic nature of the Hamiltonian plays an important role in the formation of the long-living oscillon. There have been attempts in the literature to relate the long lifetime with adiabatic invariance, but this approach just takes into account very weak nonlinearities \cite{kkt}. It was also proposed, but without a rigorous proof, that a Lyapunov exponent  governs the power law of the oscillon lifetime \cite{hc}. 
 
\begin{figure}
\begin{center}
\includegraphics[width=1 \columnwidth]{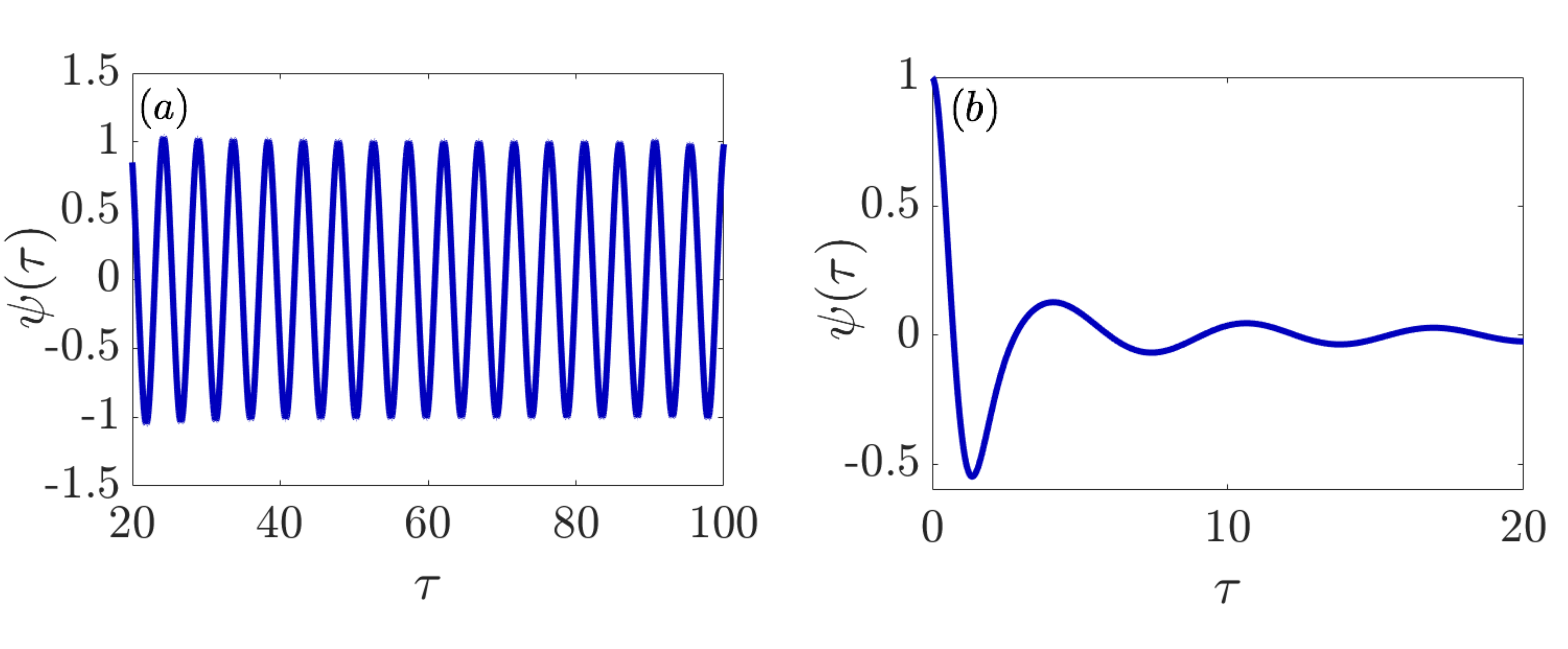}
\caption{(a) Oscillon field in time domain at $(\xi,\eta)=(0,0)$ for $A_0=1$ and $\sigma=2.86$. (b) Dispersive solution at $(\xi,\eta)=(0,0)$ for $A_0=1$ and $\sigma=1$.}
\label{fig2}
\end{center}
\end{figure}

 In figure \ref{fig1} we show the dynamics of the square modulus of the oscillon field, after subtracting the background ($A_0$), we can see after one period the original peak is fully recovered. These oscillating dynamics are quite resilient, as we can see in figure \ref{fig2}(a), even if it suffers from a weak breathing effect due to the non-integrability of the system. This is in agreement with previous results in the literature \cite{hs,cg,hc}.
 In contrast in figure \ref{fig2}(b) we can see the propagation of a dispersive solution  ($A_0 = 1$ and $\sigma = 1$). Our numerical simulations, as well as previous results \cite{hs,cg,hc}, show that the lifetime of the solution depends critically on the initial condition and it is determined by the standard deviation $\sigma$ of the gaussian ansatz. A deeper understanding of why for specific values of $\sigma$ the dynamics evolves into an oscillon, would come from a rigorous exploration of the metastability of these solutions.

\begin{figure}
\begin{center}
\includegraphics[width=1 \columnwidth]{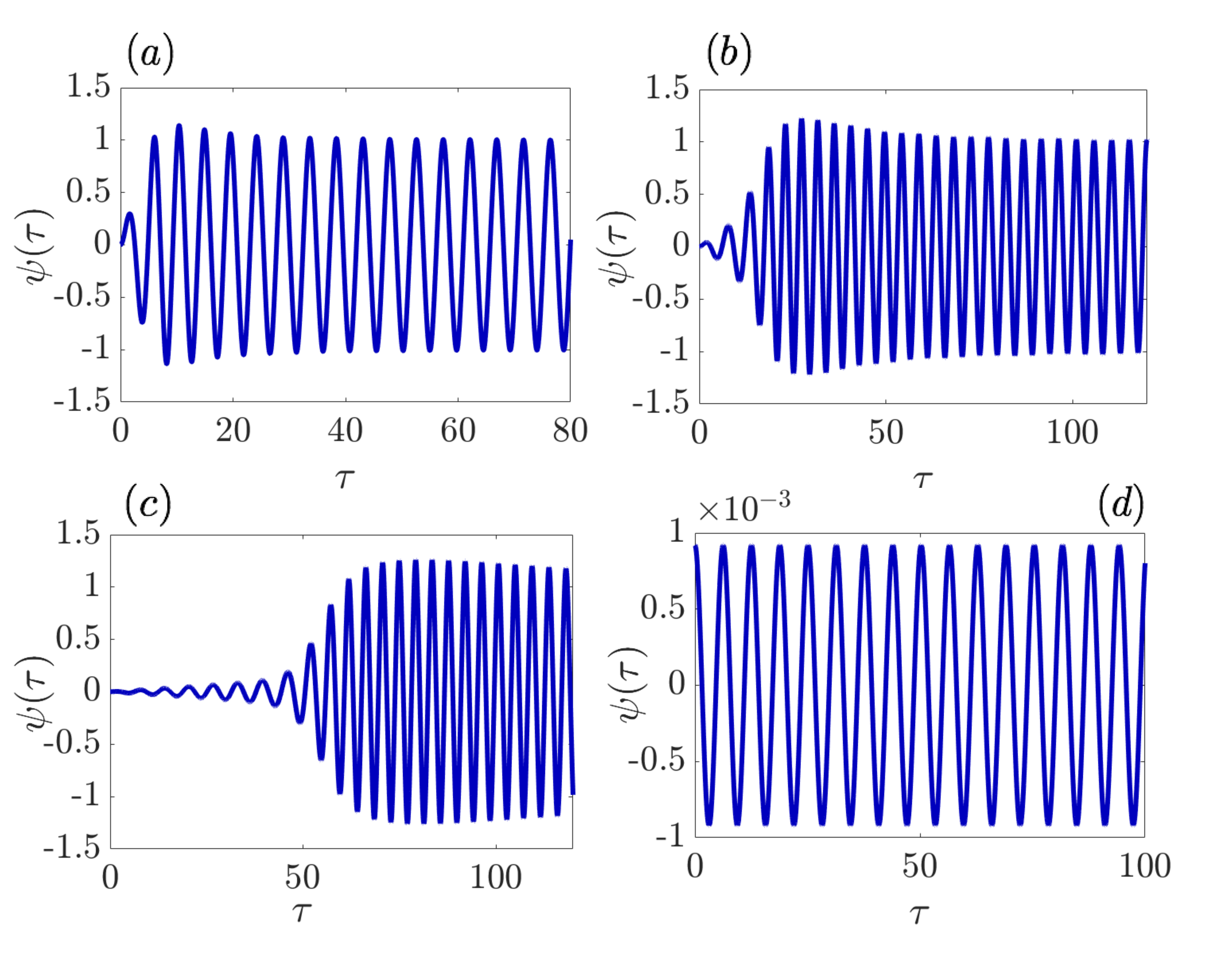}
\caption{ Early stage dynamics of the oscillon field at $(\xi,\eta)=(0,0)$ for decreasingly dispersive (increasing $r_0$ hence decreasing $\beta$) cases. (a) $r_0= 9.3$\,nm  (hence $\beta = 1$), (b) $r_0 = 29.4 $\,nm ($\beta = 10^{-1}$), (c) $r_0 = 93.2$\,nm ($\beta = 10^{-2}$), (d) $r_0 = 29.4$\,$\mu$m ($\beta = 10^{-7}$).}
\label{fig3}
\end{center}
\end{figure}

\begin{figure}
\begin{center}
\includegraphics[width=1 \columnwidth]{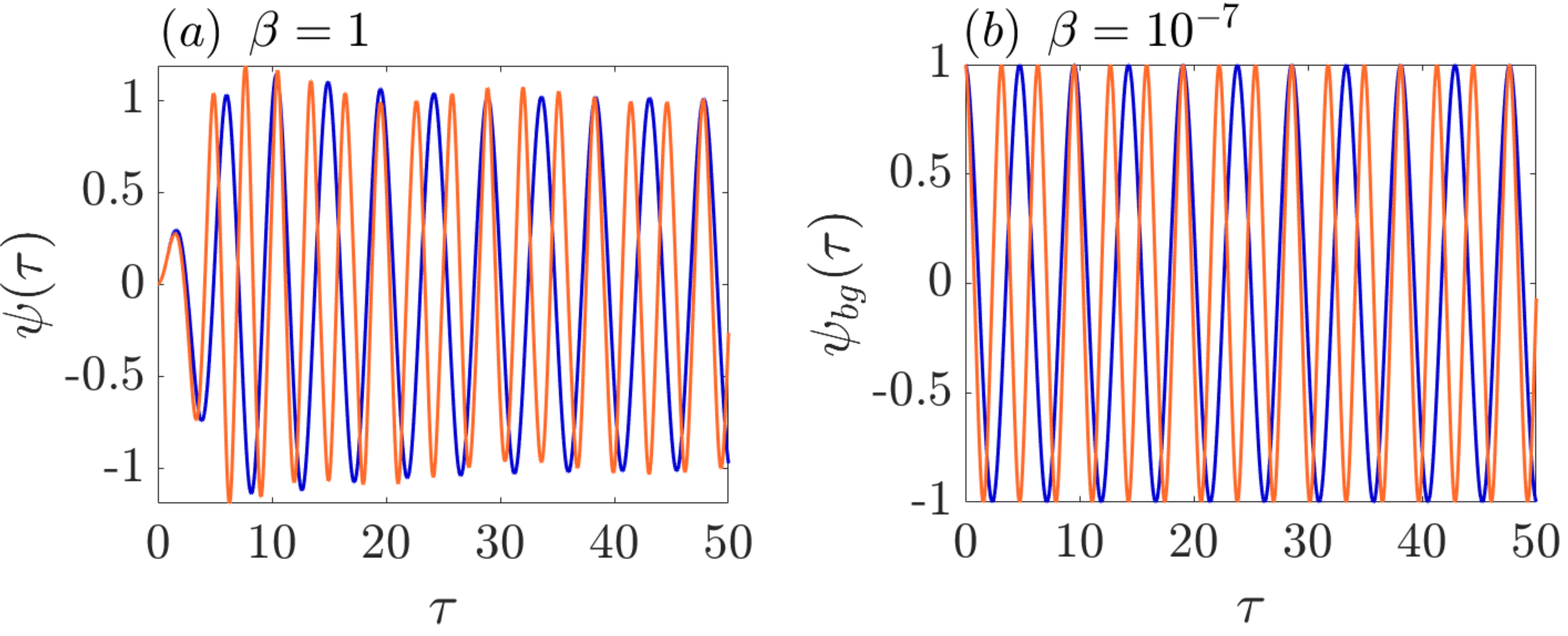}
\caption{(a) Dynamics of the oscillon  for two condensate densities $A_0 = 1$ (blue), $A_0=2$ (orange)  (b) Oscillations of the background for the dispersionless equation again for $A_0 = 1$ (blue), $A_0=2$ (orange). }
\label{fig4}
\end{center}
\end{figure}

It is interesting to study the effect of the dispersion on the dynamics of the oscillon, by varying the width of the initial gaussian hole. 
In figure \ref{fig3} we can see how the dispersion mostly affects the early stage of the dynamics. When we increase $r_0$ the dispersion term becomes less important and the formation of the oscillon is considerably delayed, as we can clearly see from figure \ref{fig3}(b) and \ref{fig3}(c). In figure \ref{fig3}(d) the dynamics completely changes, as the dispersion becomes completely negligible, and  the field at $(0,0)$ oscillates only slightly around the minimum ($\psi=0$) - note the axis scale change. The oscillations of the background instead are driven by the amplitude of the field only, as expected in the strong nonlinear limit (see figure \ref{fig4}). 

In figure \ref{fig4} we can see that the oscillon's frequency increases with the strength of the initial field, that also defines the strength of the nonlinearity. This is a well known nonlinear effect called self-phase modulation. 

To inform experimental considerations of this system, it is important to estimate some relevant physical quantities. The period of the oscillon is particularly significant because excitons are not simple bosons but composite quasi-particles with a finite lifetime and a period longer then that would be impossible to observe. This quantity can be computed from figure \ref{fig1} and the scaling relation equation (\ref{scaling}), we get $T = 20 $\,fs that is well below the exciton lifetime in two dimensional materials that is of the order of few picoseconds up to $150 $\, ps \cite{mos}.  Our model relies on the assumption of the low density limit regime, thus, an important quantity is the exciton density. From the interaction strength $U_0$ we can calculate the density for an oscillon with initial amplitude $A_0 = 1$ and radius $r_0 = 9.3 $\,nm, as $n_{ex} = 3.16 \times 10^{11}$ \,$\text{cm}^{-2}$ which, combined with the exciton Bohr radius,  gives $n_{ex} a^2_B = 0.16$ confirming the low density requirement is met. In particular for a system with a $0.2 \, $ eV gap and Coulomb coupling constant $\alpha_c  = 0.47$ we have a $1s$ state binding energy $\epsilon_b = 0.07$\,eV and  Bohr radius $a_B = 6.68$\,nm.
\section{Oscillon Interactions}

In nonlinear partial differential equations such as equation (\ref{eq10}) specific behaviours arise from the subtle interplay between nonlinearity and dispersion. In most case this is a matter of the specific numbers involved and in this section and the next we  illustrate two scenarios namely oscillon collisions and the effect of heterostructures on oscillons. The underlying physics in both cases is essentially the same as is explained as in the first sections on oscillon physics.

\begin{figure}
\begin{center}
\includegraphics[width=1 \columnwidth]{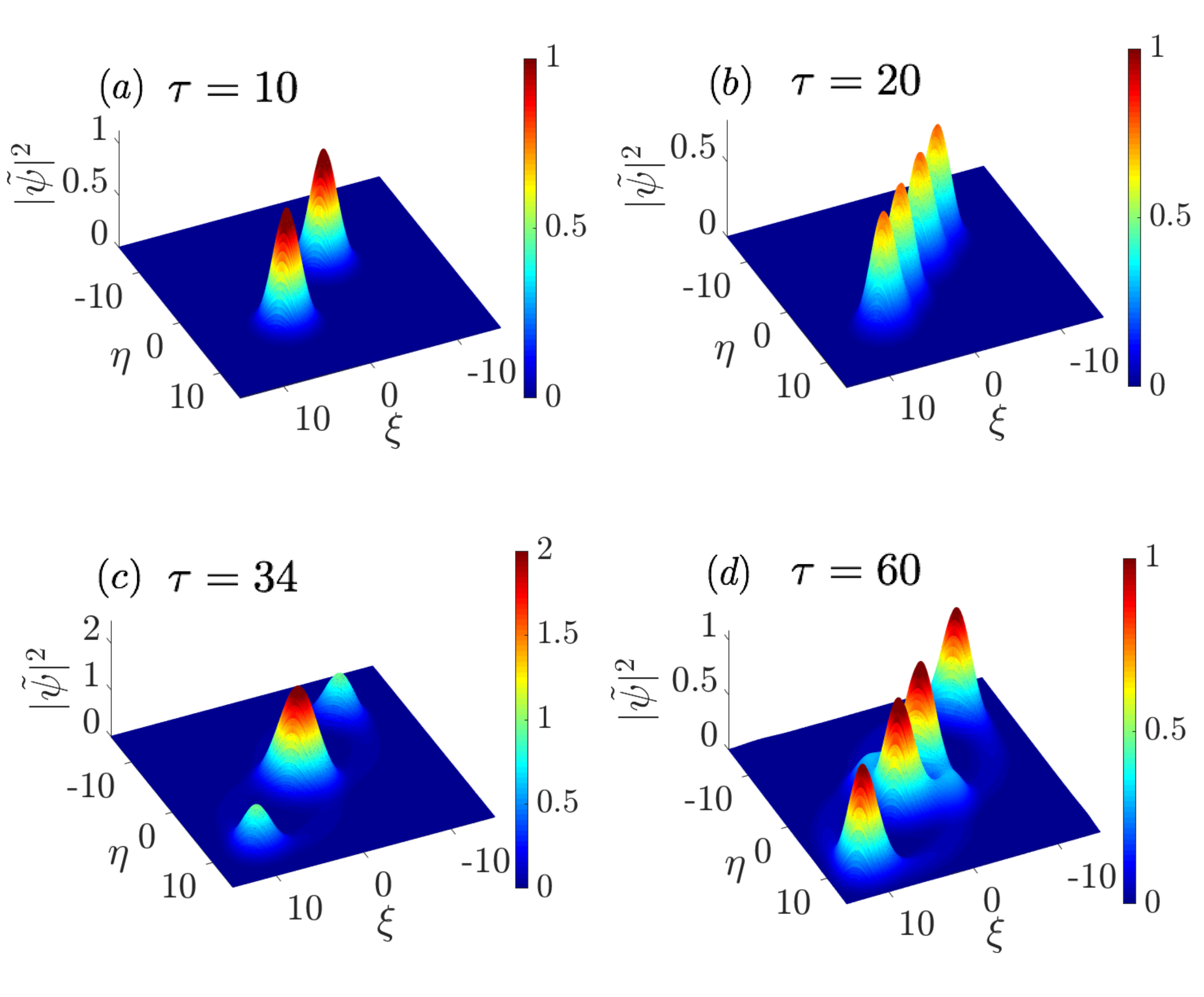}
\caption{ Scattering of two oscillons of initial amplitude $A_0 =1$, $\sigma = 2.86$ and initial speed $\tilde v_0 = 0.5$. (a) Initial state of the oscillon (b) separation of positive and negative energies, (c) collision time, (d) final state after collision. Here with $\tilde \psi$ we indicate the field after subtracting the background.}
\label{fig5}
\end{center}
\end{figure}

\begin{figure}
\begin{center}
\includegraphics[ angle=-90, width=1\columnwidth] {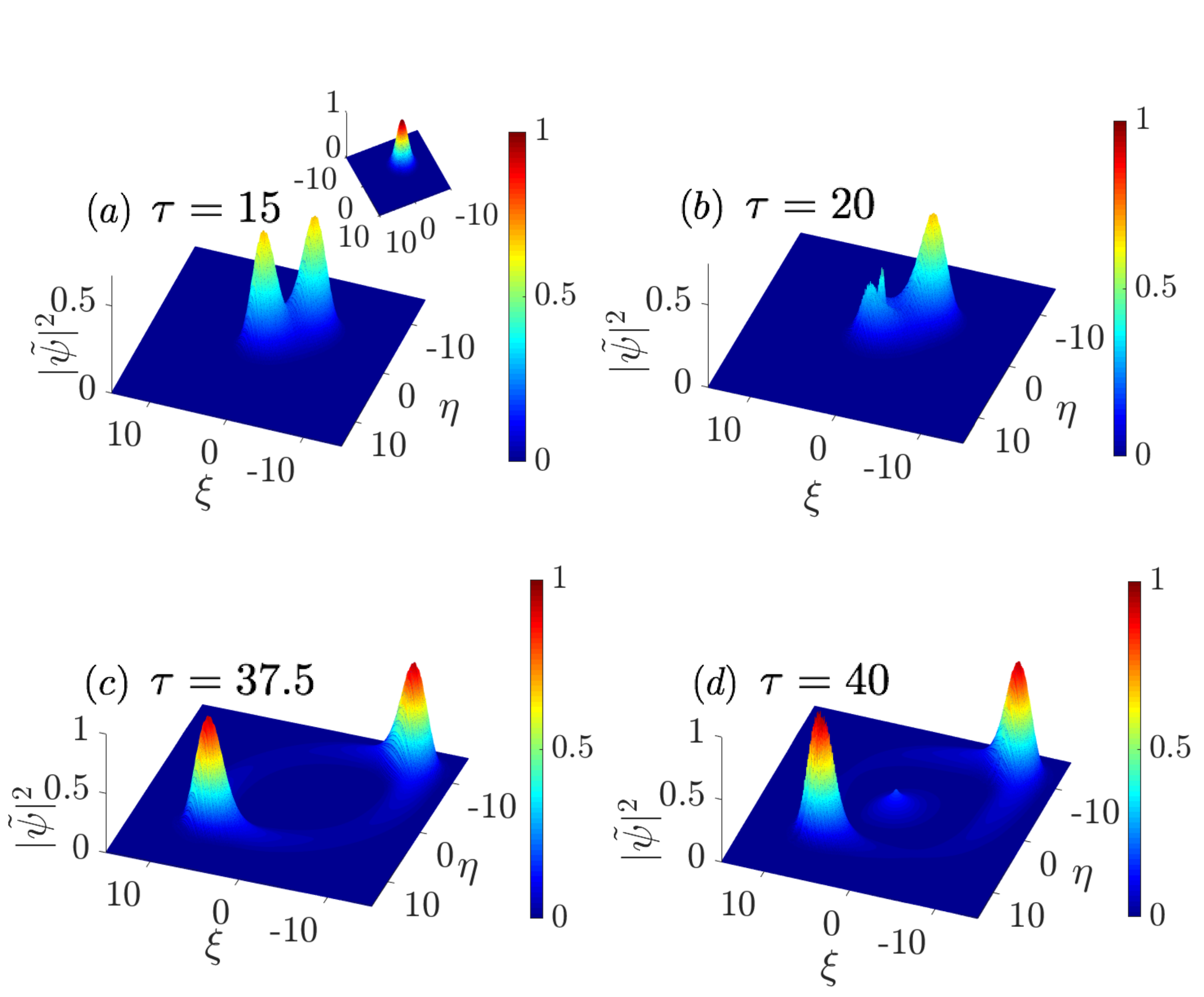}
\caption{Interaction of a single oscillon with a gaussian shaped defect located at $(\xi,\eta)=(0,0)$ and $\Gamma_0=50, \sigma_d = 10^{-2}$. (a) Early stage of the dynamics: separation of positive and negative energies. (b) Interaction with the defect, (c) and (d) late stage of the dynamics. The inset in (a)  is the initial state of the oscillon $A_0 =1$ and $\tilde v_0 = 0.5$.}
\label{fig6}
\end{center}
\end{figure}

We review first the collision of two identical oscillons \cite{hs} moving along the diagonal then we study the interaction between the oscillon and a defect. In the case of collision the initial state reads
\beq
\psi(\xi,\eta) = A_0 \left(2-  e^{-\frac{(\xi-\xi_0)^2+(\eta-\eta_0)^2}{\sigma^2}} - e^{-\frac{(\xi-\xi_1)^2+(\eta-\eta_1)^2}{\sigma^2}}\right).
\eeq
 As shown in figure \ref{fig5}  the soliton-like behaviour of the solution is preserved after the collision.
It is also interesting  to see how the oscillons interact with a defect of the condensate. This can be simulated by adding a loss term, proportional to the first time-derivative of the field. Equation (\ref{eq10}) is modified as follows 
 \beq \label{eq15}
 \partial^2_{\tau} \psi - \beta (\partial_\xi^2 + \partial_\eta^2)\psi + \psi + \psi^3 + \Gamma(\xi,\eta) \, \partial_\tau \psi = 0,
 \eeq
and we consider a gaussian defect of  the form \cite{bm} 
\beq
\Gamma(\xi,\eta) = \Gamma_0 \exp \left({-\frac{(\xi-\xi_d)^2-(\eta-\eta_d)^2}{\sigma_d^2}} \right),
\eeq
 where $\Gamma_0$ is the strength of the damping and $\{\xi_d,\eta_d\}$ the coordinates of the defect. In  figure \ref{fig6} we observe that the oscillon is resilient even to non-perturbative damping $(\Gamma_0\gg1)$.

The scaled initial speed $\tilde v_0$ we used in in this section corresponds to a velocity $v_0= 0.5\, v$. Solitons with a velocity close to the local speed of sound are known to be experimentally more stable and easier to produce in atomic non-relativistic systems. This would be a very interesting aspect to study experimentally in pseudo-relativistic materials where the role of the local speed of sound is played by the Fermi velocity $v$. Recently it has been proposed an experiment with controllable near-zero soliton velocity in atomic BEC \cite{fm}.  It would  be very interesting, ideally with a set up that is able to control the oscillon velocity, to study both the low-velocity and in principle also the ultra-relativistic limit $v_0 = v$.
\section{Condensates in heterolayers}
\begin{figure}
\begin{center}
\includegraphics[ width=1\columnwidth]{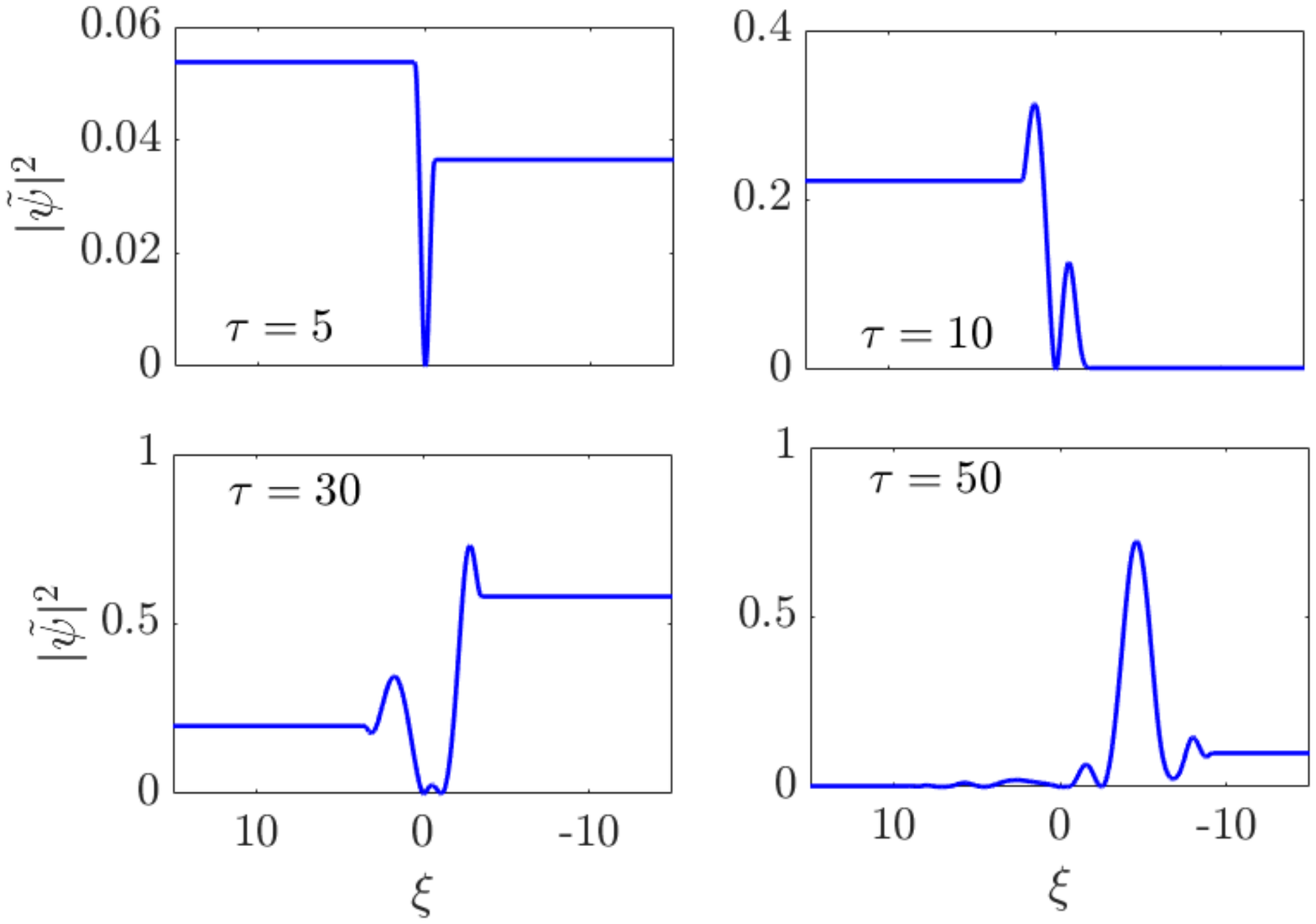}
\caption{Motion of travelling waves in two condensates with $\Delta_1 = 0.2$\,eV and $\Delta_2 = 0.3$\,eV and constant background, $\psi_0(\xi,\eta)=1$.}
\label{fig7}
\end{center}
\end{figure}

\begin{figure}
\begin{center}
\includegraphics[width=1\columnwidth]{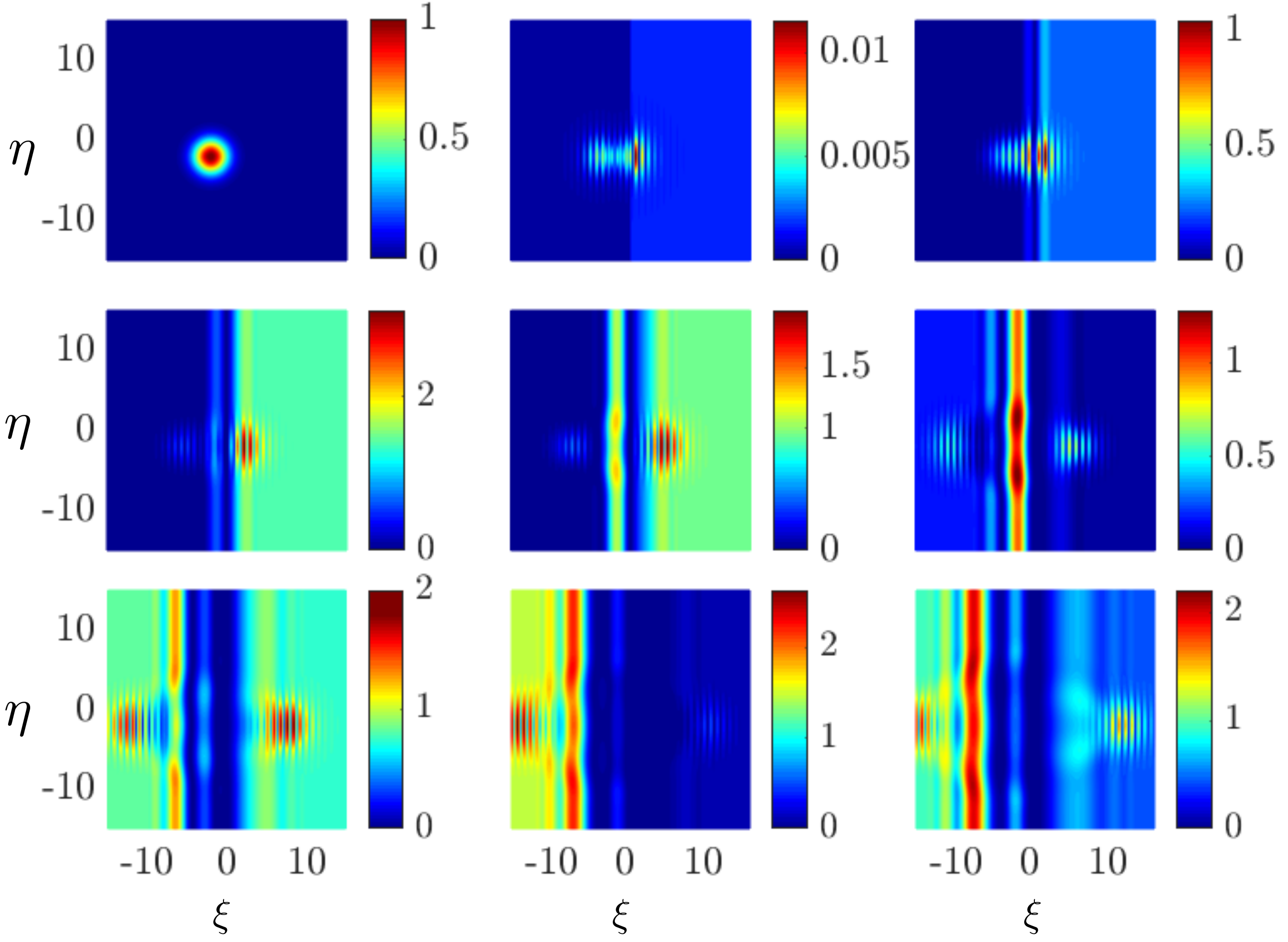}
\caption{Motion of an oscillon between two condensates with $\Delta_1 = 0.2$\,eV and $\Delta_2 = 0.3$\,eV. The initial state $\psi_0(\xi,\eta)$ is a gaussian well of the form equation (\ref{init}) located at $(\xi,\eta)=(-2,-2)$ with $A_0=1$ and $\sigma=2.86$. The sequence of snapshots, from left to right in each row, is: $\tau=0.2, 2.3, 4.4, 6.7, 9, 11.2, 13.5, 15.7, 18$. }
\label{fig8}
\end{center}
\end{figure}

\begin{figure}
\begin{center}
\includegraphics[width=1\columnwidth]{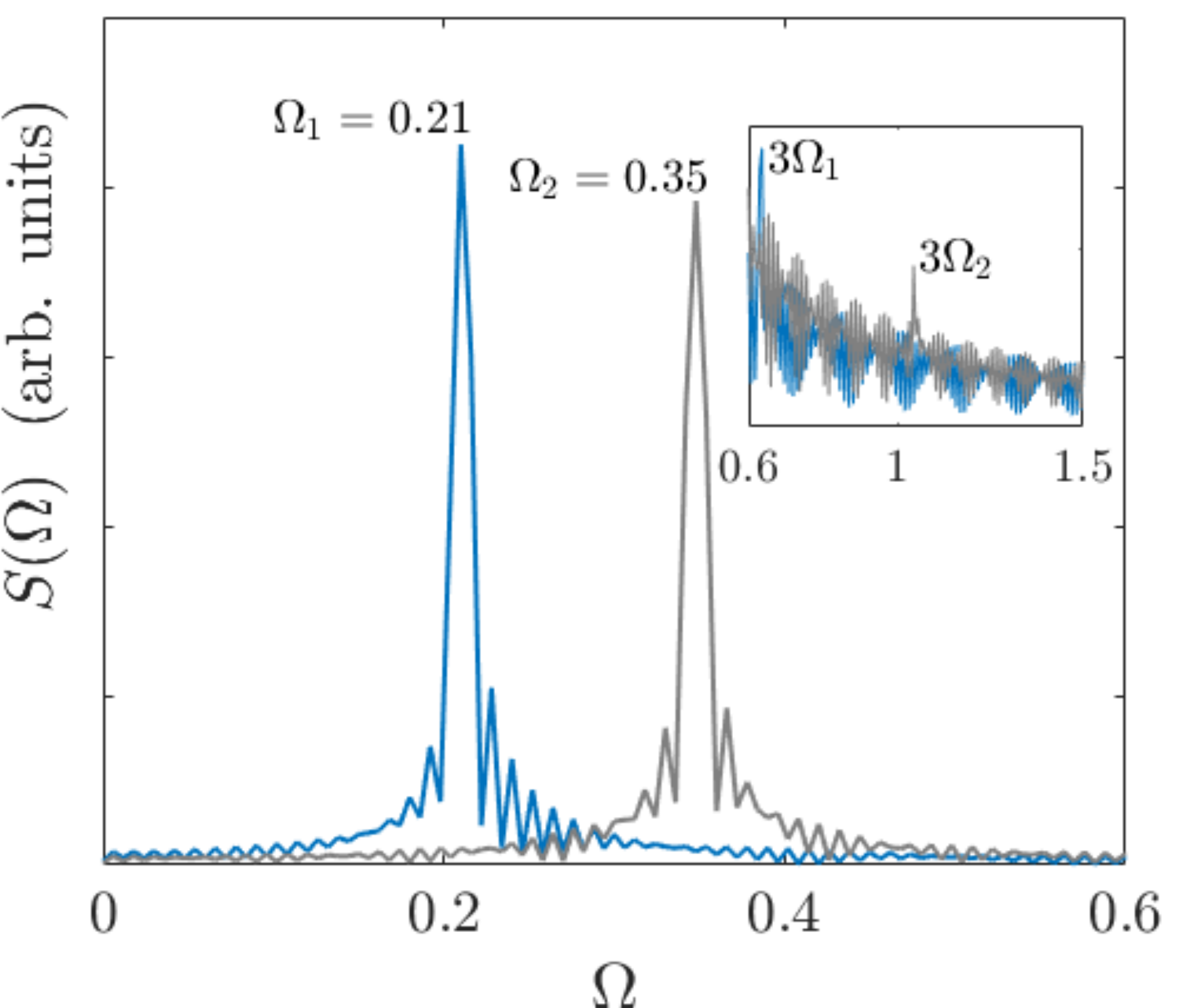}
\caption{Oscillon spectrum on both sides of the heterolayer, $\xi<0$ (blue, lower frequency peak), $\xi>0$ (grey, higher frequency peak). $\Omega$ is a dimensionless frequency related to the inverse of $\tau$. Inset: third harmonic generation.}
\label{fig9}
\end{center}
\end{figure}

We now investigate BECs spanning two connected Dirac material slabs with different energy gaps ($\Delta_1$ and $\Delta_2$) requiring an adapted version of equation (\ref{eq10}). To do so we introduce a scaled time as $\tau = t \Delta_1/\hbar$. This leads to a dimensionless equation that has the same form as equation (\ref{eq10}) on the side with energy gap $\Delta_1$ and mass $m_1 = \Delta_1/4v^2$, and that depends on the ratio between two different energy gaps $\Delta_2/\Delta_1 = m_2/m_1$ on the other side:
\beq \label{eq11}
\begin{aligned}
&\partial^2_{\tau} \psi -   \beta (\partial_\xi^2 + \partial_\eta^2)\psi + \gamma^2(\xi) \psi + \gamma(\xi) \psi^3 = 0 \\
&\psi(\xi,\eta,0) = \psi_0(\xi,\eta)
\end{aligned}
\eeq
where $\beta = 8[ \hbar v/(r_0 \, \Delta_1)]^2$ and the scaled field $\psi$ is defined as in equation (\ref{eq13}) with $\Delta = \Delta_1$. The space dependent coefficient $\gamma(\xi)$ is defined as follows
\beq
\begin{cases}
\gamma(\xi) = 1 \,\,\,\,\,\,\,~~~ \xi < 0, \\
\gamma(\xi) = \frac{\Delta_2}{\Delta_1} \,\,\,~~~ \xi > 0.
\end{cases}
\eeq
We take first the simplest case of a constant background, $\psi_0(\xi,\eta) = \psi_0$. As we can see from figure \ref{fig7}, travelling waves are generated by scattering at the boundary between the two layers and propagate throughout the two sides of the condensate. The dynamics of these waves is straightforwardly understood using a multiple scale perturbative analysis of equation (\ref{eq11}) limited to one side of the heterolayer (See Appendix C for details). 
In figure \ref{fig8} we show the motion of an oscillon located initially in the condensate with energy gap $\Delta_1$. The stability of the oscillon dynamics in not particularly influenced by the presence of the second condensate after the splitting of positive and negative energies. The oscillon moves through the second condensate and starts oscillating with a higher frequency proportional to the second gap $\Delta_2$ as we show in figure \ref{fig9}. 
From figure \ref{fig9} we can compute the frequency (in eV) of the first harmonics of the oscillons on both sides of the condensate. We get $\omega_1 = 0.041$\,eV and $\omega_2 = 0.068$\,eV. Those values are significantly lower than the two energy gap ($\Delta_1$ and $\Delta_2$) in the two sides of the heterolayer, this is because we are studying the dynamics of the system at a relatively early stage, between $\tau=10$ and $\tau = 200$, when the combined effect of the dispersion and the nonlinearity is still strong. For long time propagation the value of the frequency approaches the one of the energy gap and the system behaves like a harmonic oscillator \cite{hs}. In the inset we can observe the  generation of third harmonics in the spectrum of the bosonic field as expected from a system with third order nonlinearity. This effect is clearly not specifically related  to the heterolayer structure studied in this section and  would be present also in the spectrum of time series in figures \ref{fig2}, \ref{fig3} and \ref{fig4}.

\section{Conclusions}

In this paper we derived a simple mean field model to investigate the dynamics of Bose Einstein condensates for quasi-particle with pseudo-relativistic low energy dispersion. This approach is based on a generalisation of the Gross-Pitaevskii equation. We applied this model to the exciton dispersion of gapped Dirac material, such as doped or strained graphene and TMDs. We remark, however, that the interest of this model is not limited to these materials but could be applied to other physical systems. Magnons in TlCuCl$_3$ \cite{sl}, for example, have been proven to show a BEC phase with a relativistic dispersion relation. It could be also be generalised by including the polarisation degree of freedom, to exciton-polariton condensates in synthetic honeycomb-like photonic lattices. We studied the properties of a non-stationary, but localised solution of the model, known as an oscillon. We detailed the dynamics of the two oscillons interaction and proved that this solution is also resilient to the interaction with impurities of the background. Until now the relevance of oscillon solutions of relativistic field theories has been highlighted only in high energy physics in both 2+1 and 3+1 dimension and they have not been considered in condensed matter physics. It is important to note that the oscillon solutions we have discussed in this paper are not allowed in a simple non-relativistic BECs because the Gross-Pitaevskii  is first order in time. They are tightly related to the pseudo-relativistic dispersion of Dirac quasiparticles.  As we mentioned in the introduction oscillons of a different mathematical structure can be found in coupled non relativistic systems or in cavities with oscillating walls. Materials that show a BEC phase with pseudo-relativistic dispersion relation could represent an interesting optical analogue platform to experimentally mimic field theories.
\section{Acknowledgements} LDMV acknowledges support from the EPSRC Centre for Doctoral Training in Condensed Matter Physics (grant EP/G03673X/1) and thanks Dr. David N. Carvalho (NORDITA) for useful early discussions.
\section*{Appendix A: Wannier equation and the exciton operator}
In this appendix we give more details on the Wannier equation for excitons in pseudo-relativistic materials. Assuming the non-interacting ground state, the linear response Dirac-Bloch equation for the polarisation $P_{\k}$ is given by  \cite{sk,cm,cb,msk,lg}
\beq 
i \hbar \dot P_\k = 2( \epsilon_\k + \frac{1}2 \sum_{\k'} V_{\k\k'}) P_{\k} - \hbar \Omega^R(t),
\eeq
where $\epsilon_\k$ is the electron and hole pseudo-relativistic dispersion and $\Omega^R(t)$ is the Coulomb-renormalised Rabi frequency.  The related Wannier equation (electron-hole Coulomb problem) reads \cite{rn,sk,pk}
\beq \label{W}
2 \epsilon_\k u_{n,j} (\k) +  \sum_{\k'} V_{\k\k'} u_{n,j} (\k) = E_{n,j}  u_{n,j} (\k),
\eeq
this equation corresponds to the $k$-space representation of the Dirac-Coulomb problem \cite{rn,sk,nov,pk}
\beq \label{DC}
(2 \hbar v \boldsymbol \sigma \cdot \k + \Delta \sigma_z + V(\r) ) \vec \Psi_{nj}(\r) =  E_{n,j} \vec \Psi_{nj}(\r),
\eeq 
where $\vec \Psi_{nj}(\r)$ is a two components spinor.  Equations (\ref{W}) and (\ref{DC}) have in principle both positive and negative energy solutions. When we consider $X^0$ paraexcitons they are simply the complex conjugate of each other. The positive energy solutions of equation (\ref{W}) are given by \cite{sk}
\beq
u_j(\k) = \varphi_{j}(\k) + i \chi_{j}(\k),
\eeq
where $ \varphi_{j}(\k)$ and $ i \chi_{j}(\k)$ are the spinor components in $k$-space, where we dropped the index $n$ and we fixed the psudospin-angular momentum $j = \pm 1/2$. The exciton creation operator can be written as  
\beq
\hat c^\dagger_{\k} = \sum_{\q,j} u_{j}(\q)  \ad_{\k+\q,j} \bd_{\k-\q,-j}.
\eeq
\section*{Appendix B: The pseudo-relativistic Hamiltonian in momentum space}
In this second appendix we show how to relate the pseudo-relativistic Hamiltonian in equation (\ref{eq7a})  with the momentum space Hamiltonian in equation (\ref{eq8a}) of the main text. It is easier to deal with the interacting and non-interacting terms separately. We first rewrite the linear Hamiltonian $H_0$  
\beq \label{H0}
H_0 = \frac{1}2\int d^2r \, [ \hbar^2 \hatF_t^2 + \hbar^2 v^2 (\nabla \hatF)^2 + m^2 v^4 \hatF^2 ],
\eeq
using the expansion in terms of exciton ladder operators 
\beq \label{fi}
\hatF = \frac{1}{\sqrt{A}} \sum_\k \frac{1}{\sqrt{\mathcal E_\k}} (\hat c_\k e^{-i(\k \cdot \r - \omega_\k t)} +  \hat c^\dagger_\k e^{i(\k \cdot \r - \omega_\k t)}),
\eeq
 substituting this expression in equation (\ref{H0}), after lengthy but simple algebra \cite{das} we get 
\beq
H_0 = \sum_{\k} \mathcal E_{\k}  ~\hat c^\dag_\k \hat c_\k.
\eeq 
Let us now look at the interaction Hamiltonian 
\beq \label{normord1}
H_{I} =  U_0 \int d^2r \, \hatF^4 .
\eeq
After normal ordering, $H_{I}$ contains the following generic terms 
\beq \label{eqx2}
(\hat c_\k)^4, \,\, (\hat c^\dag_\k)^4, \, \,  \hat c^\dag_\l \hat c^\dag_\p \hat c^\dag_\q \hat c_\k, \,\, \hat c^\dag_\k \hat c_\l \hat c_\p \hat c_\q, \,\, \hat c^\dag_\k \hat c^\dag_\l \hat c_\p \hat c_\q,
\eeq
One can see that the first four terms of the list above are associated with field components [see equation~(\ref{fi})] that rotate as $e^{-4i(\k \cdot \r - \omega_\k t)}$, $e^{+4i(\k \cdot \r - \omega_\k t)}$, $e^{i(\l+\p+\q -\k)\cdot \r -i(\omega_\l + \omega_\p + \omega_\q -\omega_\k)t}$ and $e^{-i(\l+\p+\q -\k)\cdot \r +i(\omega_\l + \omega_\p + \omega_\q -\omega_\k)t}$ respectively. In these terms, due to conservation of energy and momentum, the exponents never vanish in scattering processes from a hard field potential. It is thus clear that the fifth term  is in general the dominant one since the other terms will be spatially and temporally averaged out during the evolution of the condensate. We therefore retain only the last term $\hat c^\dag_\k \hat c^\dag_\l \hat c_\p \hat c_\q$ in the normal-ordered Hamiltonian (equation~(\ref{normord1})). This procedure is standard, and is conventionally used when treating non-relativistic Hamiltonians \cite{ms}.
Considering the fifth term only we obtain the momentum conserving interaction Hamiltonian in momentum space 
\beq
H_{I}  = \frac{U_0}{2A} \sum_{\k \l \p} \frac{1}{\sqrt{\mathcal E_{\k}\mathcal E_{\l} \mathcal E_{\l + \p } \mathcal E_{\k-\p}} } \hat c^\dag_\k \hat c^\dag_\l \hat c_{\l + \p} \hat c_{\k-\p}.
\eeq 
\section*{Appendix C: Multiple scale perturbation theory for the NLKGE}
In this appendix, we introduce the multiple scale method for the nonlinear Klein-Gordon equation to show how the travelling waves propagates in a heterolayer condensate. For simplicity we limit our analysis on one side of the condensate since the dynamics is the same on both sides. 
\beq \label{eqA1}
\partial^2_{\tau} \psi -  (\partial_\xi^2 + \partial_\eta^2)\psi + m^2 \psi + \psi^3 = 0, 
\eeq
with initial conditions 
\beq
\begin{aligned}
&\psi(\xi,\eta,0) = \psi_0(\xi,\eta), \\
&\psi_\tau(\xi,\eta,0) = \psi_1(\xi,\eta),
\end{aligned}
\eeq
we set a slow space-time scale $\xi_1 = \epsilon \xi$, $\eta_1 = \epsilon \eta$, and $\tau_1=\epsilon \tau$. We can now make the following ansatz of a perturbation series for the solution
\beq \label{eqA2}
\psi(\xi,\eta,0) = \sum_n \epsilon^{n+1} \Psi_n(\xi,\xi_1,\eta,\eta_1,\tau,\tau_1,\tau_2).
\eeq
Modifying the derivatives accordingly to the slow scale transformations and inserting the ansatz, the wave equation (\ref{eqA1}), up to the second order, becomes (we will implicitly assume that initial conditions must be met at each order)
\beq
\epsilon \mathcal L_{KG} (\Psi_0) + \epsilon^2[\mathcal L_{KG} (\Psi_1) - 2 \Psi_{0,\tau_1 \tau} + 2 \Psi_{0,\xi_1 \xi} ]+ \mathcal{O}(\epsilon^3) = 0
\eeq
where $\mathcal L_{KG} = \square - m^2$ is the linear Klein-Gordon operator. \\
The order $\epsilon$ gives $\mathcal L _{KG}(\Psi_0)= 0$, that is solved by a function of the form 
\beq
\Psi_0(\xi,\xi_1,\eta,\eta_1,\tau,\tau_1) = A(\xi_1,\eta_1,\tau) e^{i(k_\xi \xi + k_\eta \eta - \omega \tau)} + c.c.
\eeq 
with the dispersion relation $\omega (\bold k)=\sqrt{\k^2 + m^2}$, $\k=(k_\xi,k_\eta)$. Proceeding  to the following order we get 
\beq
\mathcal L_{KG}(\Psi_1) = 2i(\omega A_\tau + k_\xi \partial_{\xi_1} A + k_\eta \partial_{\eta_1} A) e^{i(k_\xi \xi + k_\eta \eta - \omega \tau)} + c.c.
\eeq
this term has the same structure as the solution to the homogeneous problem, it thus represents a secularity and needs to be eliminated. This leads to the following condition on the amplitude $A(\xi_1,\eta_1,\tau)$
\beq
\omega A_\tau + k_\xi \partial_{\xi_1} A + k_\eta \partial_{\eta_1} A = 0,
\eeq
and a solution of this equation is a unidirectional travelling wave of the form $A= A(\boldsymbol \rho_1- \bold v \tau_1)$ where $\boldsymbol \rho_1 = (\xi_1,\eta_1)$ and $\bold v = (v_{\xi},v_{\eta})$ is the group velocity vector. This explains the behaviour in figure \ref{fig7} and \ref{fig8} where we see unidirectional waves propagating along the $\xi$ direction only, since the initial momentum along $\eta$ is set to zero.

\end {document}